\documentclass[twocolumn,floatfix]{revtex4}
\pdfoutput=1
\usepackage{graphicx}
\usepackage{dcolumn}
\usepackage{longtable}
\usepackage{amsmath}
\usepackage{amssymb}

\begin{document}

\title
{Highest weight irreducible representations favored by nuclear forces within SU(3)-symmetric fermionic systems}

\author
{Andriana Martinou$^1$, S. Sarantopoulou$^1$, K.E. Karakatsanis$^2$, and    Dennis Bonatsos$^1$}

\affiliation
{$^1$Institute of Nuclear and Particle Physics, National Centre for Scientific Research 
``Demokritos'', GR-15310 Aghia Paraskevi, Attiki, Greece}

\affiliation
{$^2$Department of Physics, University of Zagreb, HR-10000 Zagreb, Croatia   }

\begin{abstract}
The consequences of the attractive, short-range nucleon-nucleon (NN) interaction on the wave functions of nuclear models bearing the SU(3) symmetry are reviewed. The NN interaction favors the most symmetric spatial SU(3) irreducible representation (irrep), which corresponds to the maximal spatial overlap among the fermions.  The consideration of the highest weight (hw) irreps in nuclei and in alkali metal clusters, leads to the prediction of a prolate to oblate shape transition beyond the mid--shell region. Subsequently, the consequences of the use of the hw irreps on the binding energies and two-neutron separation energies in the rare earth region are discussed 
within the proxy-SU(3) scheme, by considering a very simple Hamiltonian, containing only the three dimensional (3D) isotropic harmonic oscillator (HO) term  and the quadrupole-quadrupole interaction. 
This Hamiltonian conserves the SU(3) symmetry and treats the nucleus as a rigid rotator.

\end{abstract}

\maketitle

\section{Most important SU(3) irreducible representations in atomic nuclei}

The SU(3 symmetry has a long history of being used in nuclear structure \cite{Kota}. Basic features of several nuclear observables are determined, often in a parameter-free way, by the irreducible representations (irreps) \cite{Wybourne,IacLie} in which the various energy bands live. In the simplest possible terms, the irreps represent the internal structure of the states bearing specific symmetry characteristics imposed by SU(3). The SU(3) irreps in nuclear physics are usually described in the Elliott notation \cite{Elliott1,Elliott2,Elliott3,Elliott4,Harvey,Wilsdon} by $(\lambda,\mu)$, with $\lambda=f_1-f_2$ and $\mu=f_2$, where $f_1$, $f_2$ are the number of boxes in the first and the second line of the relevant Young diagram \cite{Wybourne,IacLie}, exhibiting pictorially the symmetrization and antisymmetrization contents of the irrep, since boxes on the same horizontal line correspond to symmetrization, while boxes in the same vertical column represent antisymmetrization. 

One can distinguish two large families of algebraic collective models, based on their building blocks, which can be either fermions (as initially expected from the microscopic point of view, since nucleons are fermions) or bosons (under specific assumptions and approximations, allowing a system of correlated fermions to be described by bosons, as boson mapping techniques \cite{KM} indicate). 

Initially introduced by Elliott in 1958 \cite{Elliott1,Elliott2} in the sd shell, the fermionic SU(3) symmetry is based on the SU(3) symmetry of the three-dimensional (3D) isotropic harmonic oscillator (HO) \cite{Wybourne,IacLie,Smirnov}. The 3D isotropic HO SU(3) symmetry is known to be destroyed in higher shells by the spin-orbit interaction, which is the essential ingredient leading to the interpretation of the nuclear magic numbers 2, 8, 20, 28, 50, 82, 126, \dots through the shell model \cite{Mayer1,Mayer2,Jensen,MJ}. Possible extensions of the SU(3) symmetry to the shells above the sd shell have been proposed within the pseudo-SU(3) model (1973) \cite{Adler,Shimizu,pseudo1,pseudo2,DW1,DW2,Harwood,AnnArbor,Quesne,Hess,Ginocchio1,Ginocchio2}, the quasi-SU(3) model (1995) \cite{Zuker1,Zuker2,Kaneko}, and the proxy-SU(3) model (2017) \cite{proxy1,proxy2,proxy3,EPJASM,EPJPSM,EPJAHW,EPJASC}. In addition, the symplectic model (1980) \cite{Rosensteel,RW} and the Fermion Dynamical Symmetry Model (FDSM, 1987) \cite{FDSM} have been introduced. Within these models the ground state band (gsb) and possibly additional collective bands (like the $\gamma_1$ band, the first $K=4$ band, etc) are contained within a $(\lambda,\mu)$ irrep with $\mu > 0$.

SU(3) symmetry is also present within bosonic models, like the Vector Boson Model (VBM, 1972) \cite{Raychev25,Afanasev,Raychev16,RR27,Minkov1,Minkov2,Minkov3}, the Interacting Boson Model (IBM, 1975) \cite{AI,IA,IVI,FVI}, and the Interacting Vector Boson Model (IVBM, 1982) \cite{IVBM1,IVBM2}.
 Within the popular IBM, the gsb sits alone within the SU(3) irrep $(2N,0)$, where $N$ is the number of bosons corresponding to the relevant nucleus, while the $\beta_1$ and $\gamma_1$ collective bands are accommodated within the $(2N-4,2)$ irrep. 

The drastic difference between fermionic models and the IBM as far as the irrep accommodating the gsb is concerned, is a consequence of the short range nature \cite{Ring,Castenbook} of the nucleon-nucleon interaction. In a fermionic system, the overall wave function has to be antisymmetric. The short range nature of the nucleon-nucleon interaction requires the irrep for the spatial part of the wave function to be as symmetric as possible,  the spin-isospin part of the wave function characterized by the irrep conjugate to the spatial one. Looking into the mathematical details \cite{EPJAHW} one sees that this ``balance'' between the spatial and the spin-isospin part of the wave function leads in most cases to $\mu>0$.
 In the case of bosons of one type, as it is the case in the simplest version of the IBM (called IBM-1) \cite{IA}, there are no restrictions imposed by the spin-isospin part of the wave function, since on one hand the difference between protons and neutrons is ignored, thus isospin plays no role, while on the other hand the Pauli principle is absent, since we are dealing with bosons. Therefore the spatial part of the wave function can become totally symmetric without meeting any spin-isospin obstacle, thus acquiring $\mu=0$. 

Even within fermionic models, the following question concerning the irrep in which the gsb sits arises. Should this be the highest weight (hw) irrep or the irrep with the highest eigenvalue of the second order Casimir operator of SU(3), which is known to correspond to the highest eigenvalue of the operator representing the quadrupole-quadrupole interaction, being responsible for nuclear deformation? Detailed arguments presented in
Ref. \cite{EPJAHW} prove that the hw irrep is the one which should be used. In particular, the percentage of the symmetric components out of the total in a SU(3) wave function is introduced, through which it is found, that no SU(3) irrep is more symmetric than the hw irrep for a given number of valence particles in a 3D isotropic HO shell.
This choice makes no difference in the lower part of each shell, but it shows up in the upper half of each shell, having as a consequence the occurrence of a transition from prolate to oblate shapes, seen experimentally both in rare earth nuclei and in metallic atomic clusters. It is remarkable that the hw irrep provides similar predictions for the nuclear shapes within both the pseudo-SU(3) and proxy-SU(3) models, despite the significant differences in their assumptions \cite{EPJST}. A by-product of the use of the hw irrep is the theoretical justification of the dominance of prolate over oblate shapes in the ground states of even-even nuclei \cite{proxy2,EPJAHW}, a puzzle which has stayed unresolved for  a long time \cite{Hamamoto1,Hamamoto2}. 

In the remainder of this work we are going to examine the consequences of choosing the hw irrep on the calculation of binding energies and two-neutron separation energies within the proxy-SU(3) scheme.

\section{Binding energies within the proxy-SU(3) scheme}

The simplest Hamiltonian of the Elliott SU(3) Model \cite{Elliott1,Elliott2,Elliott3} is used (see Chapter 4 of Ref. \cite{Lipkin}):
\begin{equation}\label{H}
H=H_0-{\chi\over 2}QQ,
\end{equation}
where $H_0$ is the 3D isotropic HO Hamiltonian for the many-particle system:
\begin{equation}\label{H0}
H_0=\sum_{i=1}^A\left({{\bf p}_i^2\over 2m}+{1\over 2}m\omega^2{\bf r}_i^2\right)
\end{equation}
with ${\bf p}_i,{\bf r}_i,m$ being the momentum, the position and the mass of the particle, while $\omega$ is the oscillation frequency and $A$ is the mass number. If one sets $\hbar\omega=1$ in Eq. (\ref{H0}), then the eigenvalue of the $H_0$ is labeled $N_0$ and is calculated by the expression \cite{Rowe2016}:
\begin{equation}\label{N0}
N_0=\sum_{i=1}^A\left(\mathcal{N}_i+{3\over 2}\right)
\end{equation}
with $\mathcal{N}_i$ being the number of the harmonic oscillator quanta of the orbit of the $i^{th}$ particle.

 The parameter $\chi/ 2$ is the strength of the quadrupole-quadrupole ($QQ$) interaction, given by \cite{Rowe2006}:
\begin{equation}
{\chi\over 2}={\hbar\omega\over 8N_0},
\end{equation}
where $N_0$ is calculated by Eq. (\ref{N0}), while \cite{NR}:
\begin{equation}
\hbar\omega={41\over A^{1/3}} \rm{MeV}.
\end{equation}

In the Elliott SU(3) symmetry the many-quanta wave function is identified by the quantum numbers $(\lambda,\mu)$. For the nuclear ground state band, the $(\lambda,\mu)$ refer to the highest weight irreducible representations (irreps). The second order Casimir operator of the SU(3) symmetry is given by \cite{Elliott4,Harwood}:
\begin{equation}
C_2=\lambda^2+\mu^2+\lambda\mu+3(\lambda +\mu).
\end{equation}
If $L$ is the orbital angular momentum of the nuclear state, then the value of the quadrupole-quadrupole interaction in a SU(3) wave function is \cite{Harwood}:
\begin{equation}
QQ=4C_2-3L(L+1).
\end{equation}

Supposing that the depth of the 3D isotropic HO potential is $V_0$ the binding energy for a nucleus with $Z$ protons and $N$ neutrons is given by:
\begin{equation}\label{BE}
BE(Z,N)=AV_0-\left(H_0-{\chi\over 2}QQ\right).
\end{equation}
The above represents the minimum energy amount one has to offer, in order to extract the particles from the 3D-HO potential. The value of the depth of the 3D-HO potential is given approximately by \cite{NR}:
\begin{equation}
V_0\approx 50\left(1-{N-Z\over A}\right)\rm{MeV},\label{V0}
\end{equation}
but in the calculation of the binding energies with the proxy-SU(3) symmetry we have treated $V_0$ as a free parameter. The values of $V_0$ have been found using a polynomial fit $V_0=c_1+c_2N+c_3N^2$ on the data. In the rare earths this fitting procedure has given $V_0\approx 44$ MeV, which is rational in comparison with the results of Eq. (\ref{V0}).

The two proton ($S_{2p}$) or two neutron ($S_{2n}$) separation energies are given by:
\begin{gather}
S_{2p}=BE(Z,N)-BE(Z-2,N),\\
S_{2n}=BE(Z,N)-BE(Z,N-2).\label{S2n}
\end{gather}

\subsection{Calculation of $N_0$ in the harmonic oscillator shells}

The Elliott SU(3) symmetry is valid for the 3D isotropic HO shells among the magic numbers 2, 8, 20, 40, 70, 112, 168...
Each 3D-HO shell has degeneracy $(\mathcal{N}+1)(\mathcal{N}+2)$. So the first shell with $\mathcal{N}=0$ has up to 2 identical nucleons, the $\mathcal{N}=1$ shell has up to 6 identical nucleons etc. 

Suppose for instance a nucleus with 6 protons and 8 neutrons, {\it i.e. }, $^{14}$C. The first 2 protons are in an orbital with $\mathcal{N}=0$, while the rest 4 are in orbitals with $\mathcal{N}=1$. For the proton configuration:
\begin{equation}
N_{0,Z}=2\left(0+{3\over 2}\right)+4\left(1+{3\over 2}\right)=13.
\end{equation}
Similarly the 8 neutrons are distributed as follows: the first 2 have $\mathcal{N}=0$ and the last 6 are in $\mathcal{N}=1$ orbitals. So for the neutron configuration:
\begin{equation}
N_{0,N}=2\left(0+{3\over 2}\right)+6\left(1+{3\over 2}\right)=18.
\end{equation}
Finally the overall $N_0$ is given by the summation:
\begin{equation}
N_0=N_{0,Z}+N_{0,N},
\end{equation}
which means that $^{14}$C is characterized by $N_0=31$.

In a similar fashion one may reproduce the values of $N_0$ given in Table \ref{N0table} for the 3D-HO shells.

\begin{table}[htb]
\centering
\caption{$N_0$ values for the 3D isotropic HO shells.}\label{N0table}

\begin{tabular}{c c c | c c c | c c c }
\hline\noalign{\smallskip}

p & $N_0$ & $\mathcal{N}$ & p & $N_0$ & $\mathcal{N}$ & p & $N_0$ & $\mathcal{N}$ \\
\hline
 2 &   3 & 0 & 42 & 161 & 4 &  86 & 419 & 5 \\
   &     &   & 44 & 172 &   &  88 & 432 & \\
 4 &   8 & 1 & 46 & 183 &   &  90 & 445 & \\
 6 &  13 &   & 48 & 194 &   &  92 & 458 & \\
 8 &  18 &   & 50 & 205 &   &  94 & 471 & \\
   &     &   & 52 & 216 &   &  96 & 484 & \\
10 &  29 & 2 & 54 & 227 &   &  98 & 497 & \\
12 &  32 &   & 56 & 238 &   & 100 & 510 & \\
14 &  38 &   & 58 & 249 &   & 102 & 523 & \\
16 &  46 &   & 60 & 260 &   & 104 & 536 & \\
18 &  53 &   & 62 & 271 &   & 106 & 549 & \\
20 &  60 &   & 64 & 282 &   & 108 & 562 & \\
   &     &   & 66 & 293 &   & 110 & 575 & \\    
22 &  69 & 3 & 68 & 304 &   & 112 & 588 & \\
24 &  78 &   & 70 & 315 &   &     &     & \\
26 &  87 &   &    &     &   & 114 & 603 & 6 \\
28 &  96 &   & 72 & 328 & 5 & 116 & 618 & \\
30 & 105 &   & 74 & 341 &   & 118 & 633 & \\
32 & 114 &   & 76 & 354 &   & 120 & 648 & \\
34 & 123 &   & 78 & 367 &   & 122 & 663 & \\
36 & 132 &   & 80 & 380 &   & 124 & 678 & \\
38 & 141 &   & 82 & 393 &   & 126 & 693 & \\
40 & 150 &   & 84 & 406 &   &     &     & \\

\noalign{\smallskip}\hline
\end{tabular}
\end{table}

\subsection{Calculation of $N_0$ in the proxy, spin-orbit like shells}

The spin-orbit like shells extend among the spin-orbit like magic numbers 6, 14, 28, 50, 82, 126... Such shells consist by the normal parity orbitals with $\mathcal{N}$ number of quanta and by the intruder orbitals with $\mathcal{N}+1$ quanta. Within the proxy-SU(3) symmetry \cite{proxy1}, a unitary transformation \cite{EPJASM} can be applied in the intruder orbitals. This transformation reduces the number of quanta by 1 unit, which means that $\mathcal{N}+1\rightarrow \mathcal{N}$, while the last orbital, which lies highest in energy and therefore is empty for most nuclei, is ignored. Consequently the proxy, spin-orbit like magic numbers become 6-12, 14-26, 28-48, 50-80, 82-124... and each proxy, spin-orbit like shell consists by orbitals with common $\mathcal{N}_{proxy}$=1, 2, 3, 4, 5, 6, \dots quanta respectively.

As a result if a nucleus has some valence neutrons in the spin-orbit like 6-14 shell, then the value of $N_0$ is the one of the closed core $N_{0,core}$ with 6 neutrons plus the $N_{0,val}$ due to the valence neutrons in the 6-12 proxy shell with $\mathcal{N}_{proxy}=1$:
\begin{equation}
N_{0,SO}=N_{0,core}+N_{0,val}.
\end{equation}

In the example of $^{14}$C, if the 8 neutrons follow the 6-14 spin-orbit like shell then:
\begin{equation}
N_{0N,SO}=13+(8-6)\left(1+{3\over 2}\right),
\end{equation}
where the value 13 has been taken from Table \ref{N0table} for the core with 6 particles, while (8-6) gives the number of valence neutrons in the 6-12 proxy shell, which in turn consists by orbitals with $\mathcal{N}_{proxy}=1$ quanta. In general:
\begin{equation}
N_{0,SO}=N_{0,core}+\sum_{i=1}^{A_{val}}\left(\mathcal{N}_{i,proxy}+{3\over 2}\right).
\end{equation}

\subsection{Results}\label{res}

Using Eq. (\ref{BE}) we  calculated the binding energies for the rare earth nuclei within the proxy-SU(3) symmetry. In order to make clear the difference between the highest weight irreps and the irreps possessing the highest eigenvalue of the second order Casimir operator, we have made two calculations, labeled respectively by ``hw'' and ``Casimir'' in the relevant figures. The necessary irreps, obtained from Tables I and II of Ref. \cite{proxy2}, are shown in Table 2, along with the corresponding eigenvalues of the second order Casimir operator. Only the upper part of the neutron shell is shown, since in the lower part of the shell the ``hw'' and ``C'' irreps are identical and can be found in Table II of Ref. \cite{proxy2}. 
Results for the Nd isotopes are shown in Figs. 1 and 2. In the lower half of the neutron shell, in which data exist, both the ``hw'' and ``C'' claculations yield identical results, as expected. In the upper half of the shell the ``hw'' and ``C'' irreps coming out from Table I of Ref. \cite{proxy2} are considerably different, as seen in Table 2, but their influence on the binding energies is minimal, as seen in Figs. 1 and 2. 
The value of $V_0$ is the only free parameter used in this calculation. As seen in Fig. 3, $V_0\approx 44$ MeV for the Nd isotopes.

\begin{table}[htb]
\centering
\caption{Highest weight SU(3) irreps, labeled by ``hw'', and SU(3) irreps possessing the highest eigenvalue of the second order Casimir operator of Eq. (6), labeled by ``C'', are listed for the upper half of the neutron 82-126 shell, together with the corresponding eigenvalues of the second order Casimir operator of SU(3), given by Eq. (6).  See subsection \ref{res} for further discussion. 
}\label{irrtable}

\begin{tabular}{c l l r r  }
\hline\noalign{\smallskip}

    & irrep & irrep & $C_2$& $C_2$\\
 N  & hw    &   C   &  hw  &  C   \\ 
\hline
104 & 54,12 & 20,44 & 3906 & 3408 \\
106 & 50,16 & 26,40 & 3754 & 3514 \\
108 & 48,16 & 28,38 & 3520 & 3490 \\
110 & 48,12 & 26,38 & 3204 & 3300 \\
112 & 50,4  & 20,40 & 2878 & 2980 \\
114 & 40,14 & 24,34 & 2518 & 2722 \\
116 & 32,20 & 24,30 & 1900 & 2358 \\
118 & 26,22 & 20,28 & 1876 & 1888 \\
120 & 22,20 & 22,20 & 1450 & 1450 \\
122 & 20,14 & 20,14 &  978 &  978 \\
124 & 20,4  & 20,4  &  568 &  568 \\

\noalign{\smallskip}\hline
\end{tabular}
\end{table}

\begin{figure}
\centering
\includegraphics[width=85mm]{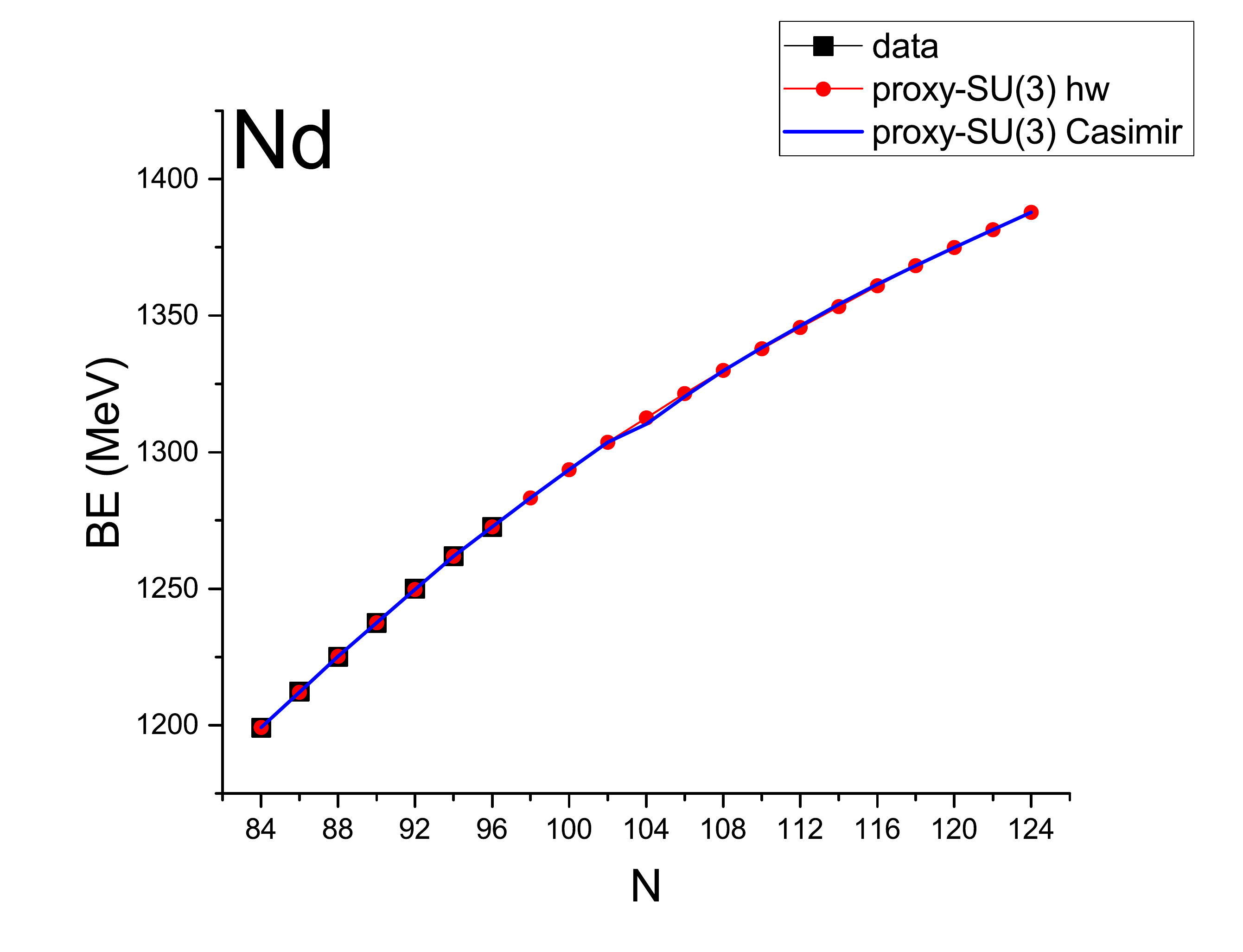}
\caption{The predictions for the binding energies of the Nd isotopes within the proxy-SU(3) symmetry, calculated with the highest weight SU(3) irrep, labeled by ''hw'', or with the SU(3) irrep possessing the highest eigenvalue of the second order Casimir operator given by Eq. (6), labeled by ``Casimir''. The data have been taken from the Atomic Mass Evaluation 2016 \cite{Wang2017}. See subsection \ref{res} for further discussion. }
\label{BENd}
\end{figure}
\begin{figure}
\centering
\includegraphics[width=85mm]{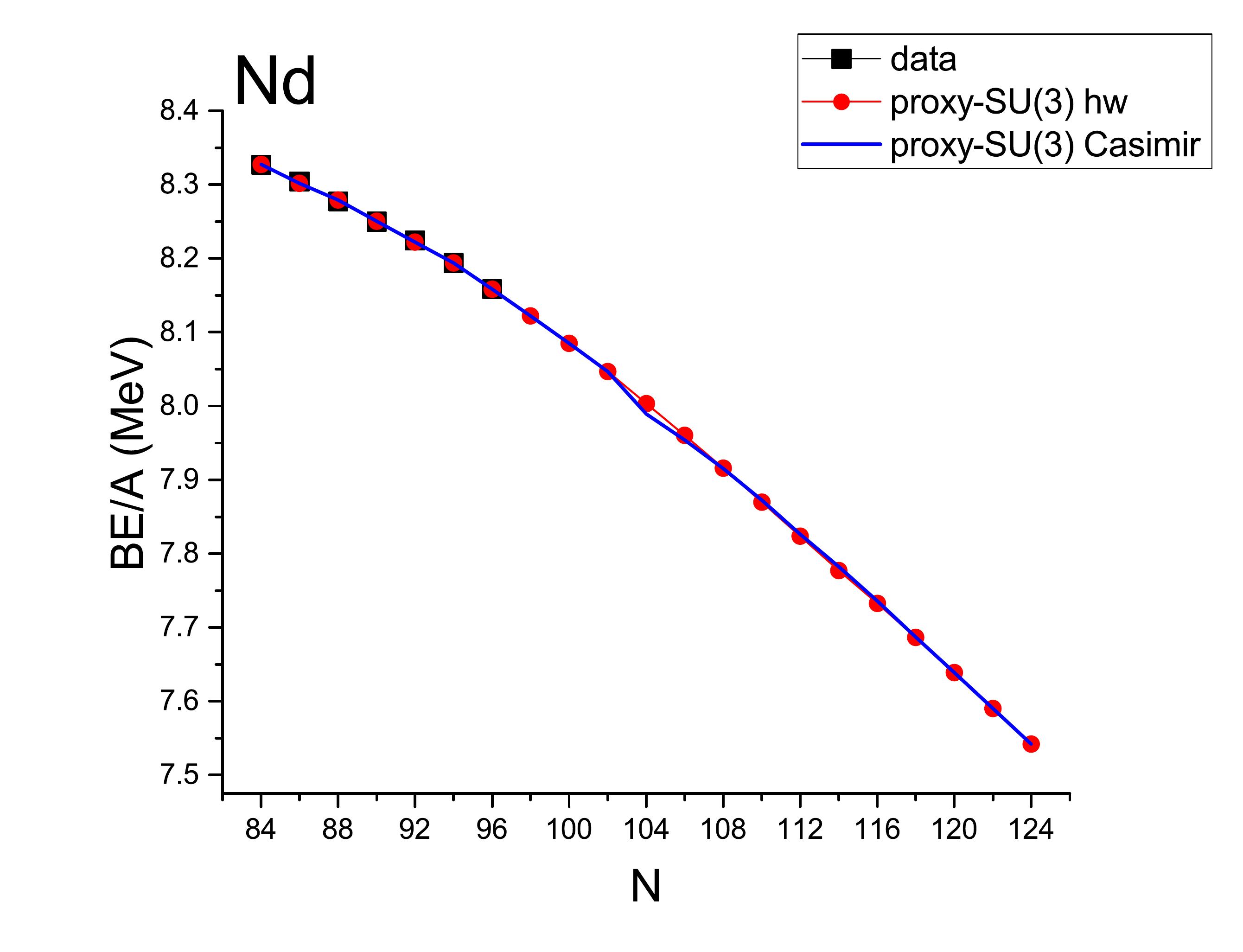}
\caption{The predictions for the binding energies $BE$ over the mass number $A$ for the Nd isotopes within the proxy-SU(3) symmetry, calculated with the highest weight SU(3) irrep, labeled by ''hw'', or with the SU(3) irrep possessing the highest eigenvalue of the second order Casimir operator given by Eq. (6), labeled by ``Casimir''. 
The data have been taken from the Atomic Mass Evaluation 2016 \cite{Wang2017}.  See subsection \ref{res} for further discussion. }
\label{BEANd}
\end{figure}
\begin{figure}
\centering
\includegraphics[width=85mm]{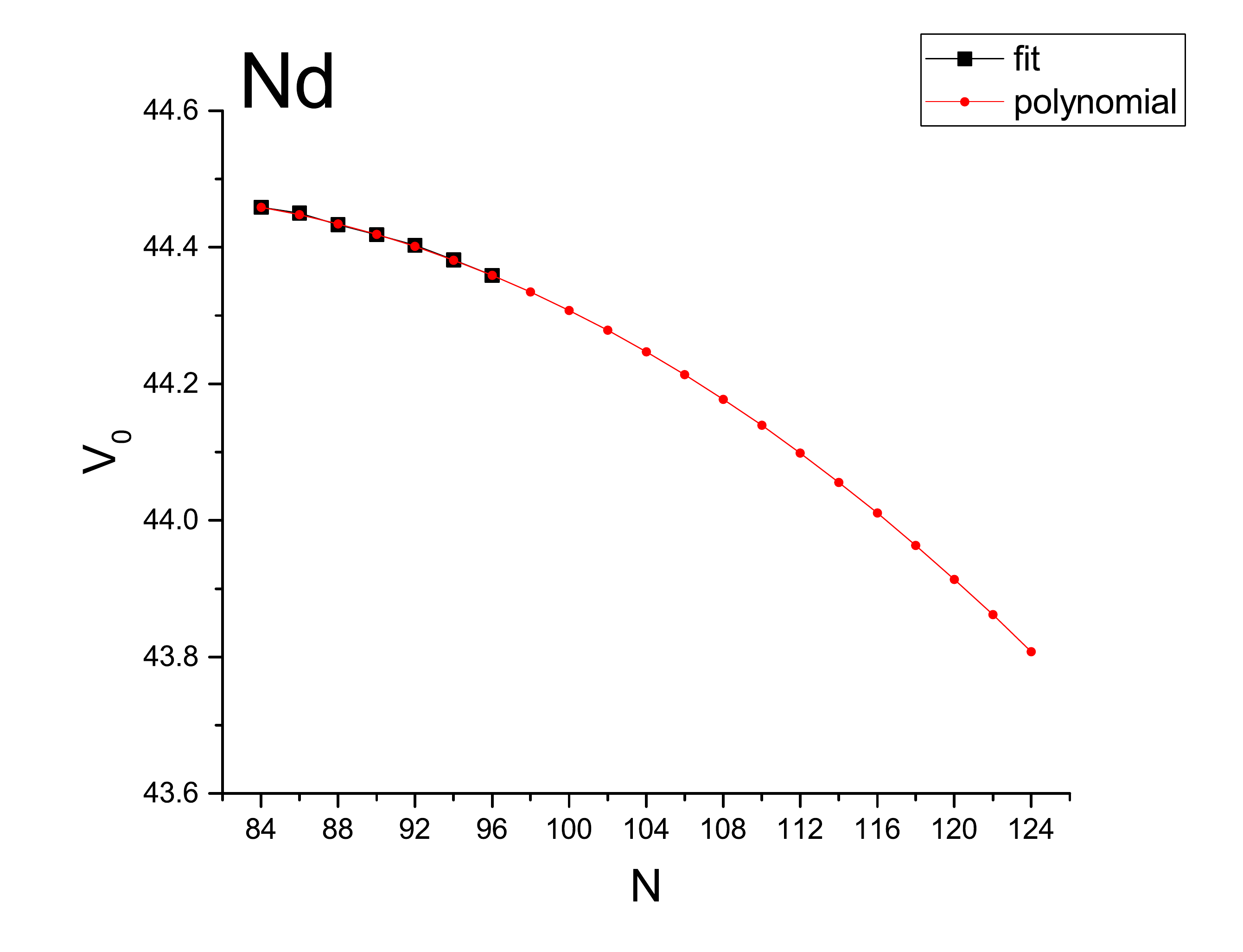}
\caption{The fitted values of $V_0$ for the Nd isotopes. A polynomial fit $V_0=c_1+c_2N+c_3N^2$ has been used. The values of the constants $c_1=42.86274$, $c_2=0.04289$, $c_3=-2.84426\cdot 10^{-4}$ have been calculated using the \textit{Origin} code. }
\label{V0Nd}
\end{figure}
\begin{figure}
\centering
\includegraphics[width=85mm]{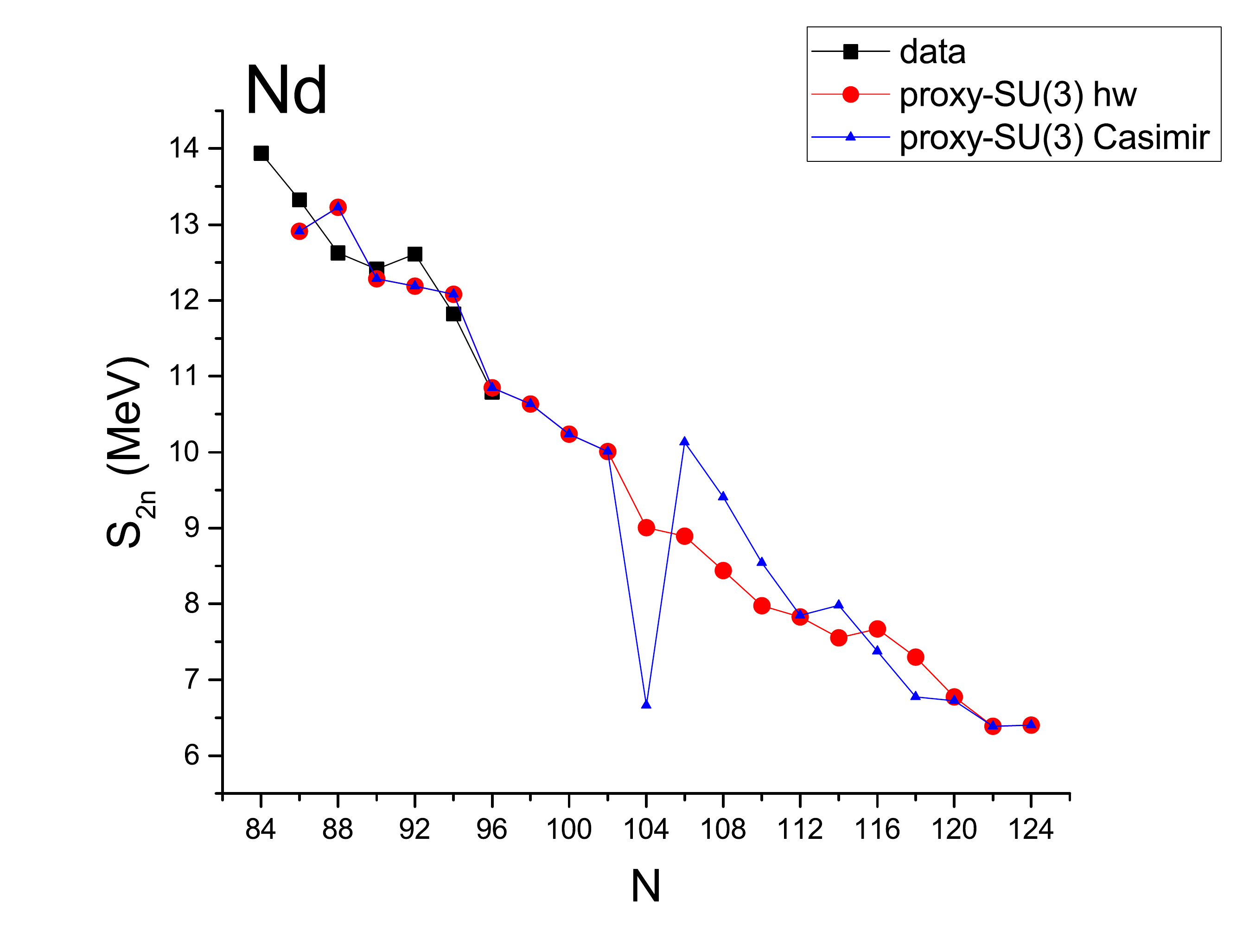}
\caption{The two neutron separation energies for the Nd isotopes. Eq. (\ref{S2n}) has been used for the calculation. Only the depth $V_0$ has been used as a free parameter, which has been fitted to data. Whenever no data exist, a linear fit has been used (see Fig. \ref{V0Nd}). Calculations have been carried out both with the highest weight SU(3) irrep, labeled by ''hw'', and with the SU(3) irrep possessing the highest eigenvalue of the second order Casimir operator given by Eq. (6), labeled by ``Casimir''. See subsection \ref{res} for further discussion.  }
\label{S2nNd}
\end{figure}

$S_{2n}$ have been calculated using Eq. (\ref{S2n}), the results for the Nd isotopes being shown in Fig. 4. The ``hw'' proxy-SU(3) predictions for the $S_{2n}$ exhibit a staircase behavior, which is similar to the one exhibited by the data, while the ``C'' predictions in the upper half of the neutron shell are considerably different, especially immediately above the midshell, where the Casimir eigenvalues of the ``hw'' and ``C'' cases differ most, as seen in Table 2.  It is encouraging that good coincidence with the data is achieved in the ``hw'' case with the use of the simplest SU(3) Hamiltonian of Eq. (\ref{H}).
 The inclusion of the spin-orbit interaction \cite{Elliott4} will mix the highest weight SU(3) irrep with other SU(3) irreps (see Eq. 4.13b of Ref. \cite{Wilsdon}), thus we expect that the spin-orbit interaction will soften this staircase behavior. It would be interesting to extend the present calculations to higher $Z$, in which data exist in the upper half of the neutron shell, where the differences between the ``hw'' and ``C'' irreps are large, in order to see which of the two approaches leads to better agreement to the data. Work in this direction is in progress. 

\section*{Acknowledgements}

This research is co-financed by Greece and the European Union (European Social Fund- ESF) through the Operational Programme ``Human Resources Development, Education and Lifelong Learning 2014-2020" in the context of the project ``Nucleon Separation Energies" (MIS 5047793).

\end{document}